\documentclass[conference,10pt]{IEEEtran}
\IEEEoverridecommandlockouts
\usepackage{cite}
\usepackage{amsmath,amssymb,amsfonts}
\usepackage{algorithmic}
\usepackage{graphicx}
\usepackage{textcomp}
\usepackage{hyperref}
\usepackage{braket}
\usepackage{bm}
\usepackage{bbold}
\usepackage{url}

\usepackage{ulem}
\usepackage[dvipsnames]{xcolor}

\def\BibTeX{{\rm B\kern-.05em{\sc i\kern-.025em b}\kern-.08em
    T\kern-.1667em\lower.7ex\hbox{E}\kern-.125emX}}
\begin{document}

\title{Iterative warm-start optimization with quantum imaginary time evolution
\thanks{
 This material is based upon work supported by the U.S. Department of Energy, Office of Science, Office of Advanced Scientific Computing Research under field work proposal ERKJ445, 
 Award Number~89243024SSC000129. 
 This research used resources of the Compute and Data Environment for Science (CADES) at the Oak Ridge National Laboratory, which is supported by the Office of Science of the U.S. Department of Energy under Contract No. DE-AC05-00OR22725.  This manuscript has been authored by UT-Battelle, LLC, under contract DE-AC05-00OR22725 with the US Department of Energy (DOE). 
SH 
was supported under Prime Contract No. 80ARC020D0010.
 The US government retains and the publisher, by accepting the article for publication, acknowledges that the US government retains a nonexclusive, paid-up, irrevocable, worldwide license to publish or reproduce the published form of this manuscript, or allow others to do so, for US government purposes. DOE will provide public access to these results of federally sponsored research in accordance with the DOE Public Access Plan (\url{https://www.energy.gov/doe-public-access-plan}}
}

\author{\IEEEauthorblockN{Phillip C. Lotshaw\IEEEauthorrefmark{1}, Titus Morris\IEEEauthorrefmark{2}, Stuart Hadfield\IEEEauthorrefmark{3}, and Ryan Bennink\IEEEauthorrefmark{4}}
\IEEEauthorblockA{\IEEEauthorrefmark{1}Quantum Information Science Section\\
Oak Ridge National Laboratory, Oak Ridge, 37830, USA\\
Email: lotshawpc@ornl.gov}
\IEEEauthorblockA{\IEEEauthorrefmark{2}Quantum Information Science Section\\
Oak Ridge National Laboratory, Oak Ridge, 37830, USA\\
Email: morristd@ornl.gov}
\IEEEauthorblockA{\IEEEauthorrefmark{3}Quantum Artificial Intelligence Laboratory,\\
USRA RIACS,\\
NASA Ames Research Center, Moffett Field, 94035, USA\\
Email: stuart.hadfield@nasa.gov}
\IEEEauthorblockA{\IEEEauthorrefmark{4}Quantum Information Science Section\\
Oak Ridge National Laboratory, Oak Ridge, 37830, USA\\
Email: benninkrs@ornl.gov}

}

\maketitle

\begin{abstract}
Approximate combinatorial optimization is a promising use case for quantum computers.  The quantum optimization algorithms often employ a fixed ansatz that evolves an unbiased initial state towards states with better values of the optimand, then samples the states to determine an approximately optimal solution.  However, promising alternative approaches have considered ``warm-start" and sampling-based methods that instead begin from the best known solution, which can be directly optimized with the quantum computer and updated as new information becomes available, potentially outperforming the fixed ans\"atze.  Here we use these ideas to design a nonvariational quantum algorithm for combinatorial optimization. At each step the algorithm begins with a state superposed around the best known solution, then drives it to lower energy using quantum imaginary time evolution.  These nonvariational, initial-state-dependent circuits are determined using analytic equations that are evaluated using only a conventional computer. After implementing the circuits, the state is sampled, potentially obtaining a new best-known solution to use as the initial state at the next iteration. Using simulations of the algorithm solving MaxCut on 3-regular graphs with 30 or fewer vertices and a shot budget of 100 total shots, the approach obtains median solutions within 95\% of the global optimum and finds optimal solutions in 11\% or more of cases, significantly outperforming random and simplified classical search procedures.  We discuss several future directions.
\end{abstract}


\section{Introduction}

Combinatorial optimization poses significant challenges in applications throughout society.  These problems often belong to the NP-hard computational complexity class, preventing exact solutions from being obtained for large instances due to resource requirements that grow exponentially with the problem size.  Heuristic approaches with polynomial resource requirements are often employed instead, to obtain approximate solutions with fewer resources, but with limited performance.  

Quantum algorithms have been proposed as alternative approaches for solving combinatorial optimization problems using entirely different principles, providing new routes to potentially obtain faster or better approximate solutions. Significant research efforts have studied quantum annealing \cite{kadowaki1998quantum,farhi2001quantum,albash2018adiabatic} and the quantum approximate optimization algorithm (QAOA) \cite{farhi2014quantum,hadfield2019quantum,blekos2024review,abbas2024challenges} as generic optimization approaches.  However, there are limitations to these approaches, in terms of long run times for adiabatic evolution and deep circuits for QAOA. Parameter optimization for variational algorithms 
is itself a challenging problem~\cite{bittel2021training}; while this difficulty has been somewhat addressed for QAOA through parameter transfer heuristics \cite{lotshaw2021empirical, shaydulin2023parameter, lotshaw2023approximate, braydwood2026qaoa, sureshbabu2024parameter, galda2023similarity, galda2021transferability, brandao2018fixed} or other approaches \cite{zhou2020quantum ,sud2024parameter,magann2022feedback, brady2025feedback,guo2026setting}, it is much less clear how to efficiently optimize parameters in variants of QAOA \cite{herrman2022multi, gaidai2024performance, kazi2025analyzing, vijendran2024expressive, goldstein2024convergence, hadfield2019quantum, chandarana2022digitized} and alternative variational algorithms \cite{morris2024performant,wang2023symmetry ,wang2025imaginary ,chai2025optimizing}. We attempt to address these issues with a nonvariational, low-depth algorithm.

Recently, some of us contributed to an alternative approach to quantum combinatorial optimization based on quantum imaginary time evolution \cite{morris2024performant}; related studies are in Refs.~\cite{wang2023symmetry ,wang2025imaginary ,chai2025optimizing,alam2023solving}.  The approach used shallow parameterized circuits that were updated based on measurements from the quantum device, and it was shown to consistently converge to optimal solutions for a variety of instances of the Max-Cut problem, in simulations and on hardware. Despite the empirical success of the approach, it nonetheless required a large number of queries to the quantum device for circuit optimization.  Here we instead use iterative updates of a product initial state, obtaining approximate solutions with far fewer queries to the quantum device. 

The current approach combines aspects of Ref.~\cite{morris2024performant} with ``warm-start" and sampling ideas related to previous research concerning warm-started quantum approximate optimization algorithms \cite{egger2021warm,wurtz2021classically,tate2023bridging,tate2023warm,goldstein2024convergence,tate2025warm,feeney2025better,lopez2025non,yuan2025iterative,bucher2026constrained, maciejewski2026quantum}, quantum enhanced simulated annealing \cite{marshall2026quantum} or Markov-chain Monte Carlo \cite{layden2023quantum}, and noise-directed adaptive remapping \cite{maciejewski2025improving,maciejewski2026quantum,hadfield2026noise}, as well as related ideas 
previously explored in the context of 
quantum annealing~ \cite{duan2013alternative,chancellor2017modernizing,karimi2017effective,grass2019quantum,venturelli2019reverse}. Our approach iteratively updates the incumbent (best known) solution using circuits that are initial-state dependent, shallow, and nonvariational.  Simulations of these circuits identify high quality solutions for instances of the Max-Cut problem using a modest budget of 100 shots per instance.  The basic algorithmic structure is amenable to a variety of changes that can potentially improve its performance, for example, by using different quantum evolutions or initializations.  Overall the approach provides a new quantum optimization algorithm that functions with shallow circuits and minimal numbers of queries to the quantum device.

\section{Iterative quantum combinatorial optimization}

We consider optimization of a function $C(\bm z)$ with an $n$-bit string argument $\bm z = (z_1,\ldots,z_n)$ with $z_i \in \{0,1\}$.  For optimization on the quantum computer, each bitstring corresponds to a computational basis state $\ket{\bm z}$ and the function is encoded in the eigenspectrum of a Hamiltonian operator
\begin{equation} H\ket{\bm z} = C(\bm z)\ket{\bm z}.\end{equation}
We focus on the unweighted maximum cut problem, which is a common benchmarking problem for quantum algorithms.  An instance of the problem is defined in reference to a graph $G = (V,E)$ with vertex set $V$ and edge set $E$, and the goal is to assign bit values $z_i$ to each vertex $i$ such that the number of edges connecting vertices in different sets is maximized.  For this problem the Hamiltonian takes the simple form
\begin{equation} \label{H} H = \sum_{(i,j) \in E} Z_i Z_j. \end{equation}
Optimal solutions $\ket{\bm z^\text{opt}}$ are eigenstates of $H$ with the minimum eigenvalue $C^\text{min}=C(\bm z^\text{opt})$.  Suboptimal solutions are eigenstates with larger eigenvalues. 

\begin{figure}
\centering
\includegraphics[width=6cm,height=6cm,keepaspectratio]{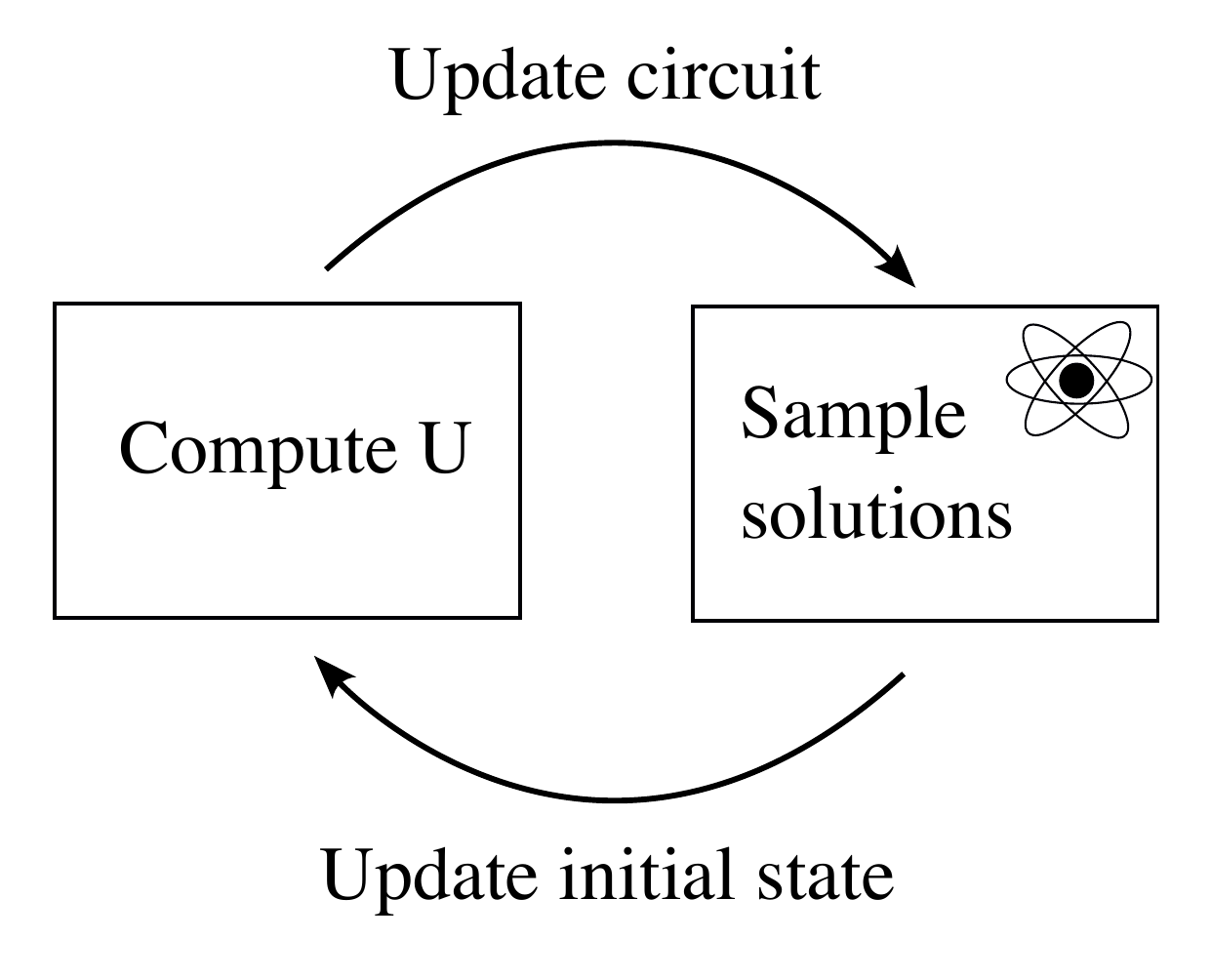}
\caption{The iterative quantum optimization algorithm uses a conventional computer (left) to compute a circuit that attempts to decrease the energy of the best-known solution, then sends the circuit to a quantum computer (right) which executes the circuit to update the best-known solution.}
\label{schematic}
\end{figure}

We designed a quantum algorithm to approximately solve these problems, following steps shown schematically in Fig.~\ref{schematic}. It begins with a quantum product state superposed around a best known solution.  For the state, we numerically compute a circuit that lowers its energy through quantum imaginary time evolution at leading order, using analytic equations that are solved nonvariationally on a conventional computer.  Applying the circuit to the state and sampling produces a pool of candidate solutions.  If a better solution is obtained, then the state and circuit are updated, else the procedure is repeated.  The algorithm terminates after a fixed number of iterations, returning the best known solution.  Overall the algorithm progresses from an initial solution to solutions that can only improve in quality, hopefully progressing toward a high-quality final result. 

\section{Algorithm details} 

We now present a detailed description of each step of the algorithm.  At a given iteration, we begin with an initial state $\ket{\psi^0}$ derived from the best-known current solution $\ket{\bm z^\text{best}} = \ket{z_1^\text{best},z_2^\text{best},\ldots,z_N^\text{best}}$. The goal is to drive this state to a lower energy using a state-dependent unitary that can be determined without requiring samples from the quantum device.  

A variety of different unitary evolutions can be chosen, for example, variations of QAOA \cite{egger2021warm,wurtz2021classically,tate2023bridging,tate2023warm,goldstein2024convergence,tate2025warm,feeney2025better,lopez2025non,yuan2025iterative,marshall2026quantum,maciejewski2026quantum,bucher2026constrained} or Trotterized time evolution circuits \cite{layden2023quantum}.  Here we consider an approach based on quantum imaginary time evolution, which has shown a variety of successes in approximately solving combinatorial problems \cite{morris2024performant,wang2023symmetry ,wang2025imaginary ,chai2025optimizing,alam2023solving}.  This approach implements a unitary that approximates the imaginary time evolution
\begin{equation} \label{QITE} \ket{\psi^\text{QITE}} \approx \ket{\psi^\text{ITE}}\end{equation}
where 
\begin{equation} \ket{\psi^\text{QITE}} = U^\text{QITE}\ket{\psi^0} \end{equation} 
is generated from a unitary quantum imaginary time evolution circuit while 
\begin{equation} \ket{\psi^\text{ITE}} = \frac{e^{-H\tau} \ket{\psi^0}}{\sqrt{ \bra{\psi^0}e^{-2H\tau}\ket{\psi^0} }}. \end{equation}
is the corresponding conventional imaginary time evolution of the same starting state. The imaginary time evolution exponentially increases the probability of optimal solutions in the support of $\ket{\psi^0}$; in principle this can solve any NP-hard optimization problem, so it cannot be done efficiently in general (unless NP-hard optimization problems can be solved in polynomial time). It is also non-unitary, so requires a unitary approximation $U^\text{QITE}$ to be implemented on a quantum computer via $\ket{\psi^\text{QITE}}$.  This unitary can be determined using classical processing informed by quantum state characterization, typically based on measurements of the state \cite{morris2024performant,wang2023symmetry ,wang2025imaginary,chai2025optimizing,alam2023solving,motta2020determining}.  The current approach determines $U^\text{QITE}$ using only classical processing, while quantum state characterization is performed analytically for a simple class of product initial states, as we now describe.

The candidate bitstring $\ket{\bm z^\text{best}}$, on its own, is not a suitable initial state for imaginary time evolution as in (\ref{QITE}).  The reason is that it is an eigenstate of $H$, which is a fixed point of the imaginary time evolution.  To formulate a circuit amenable to imaginary time evolution, we therefore begin by applying a unitary $U^\text{mix}$ to $\ket{\bm z^\text{best}} $ to generate a superposed state $\ket{\psi^0} = U^\text{mix} \ket{\bm z^\text{best}}$ that has an overlap with $\ket{\bm z^\text{best}}$ that is at least as large as the overlap with any other computational basis state. We choose $U^\text{mix}$ as a product of Pauli-$Y$ rotations, such that
\begin{equation} \label{superposed state} \ket{\psi^0} = \bigotimes_i [\cos(\varphi) \ket{z_i^\text{best}} + \sin(\varphi) \ket{1-z_i^\text{best}}] \end{equation}
In this equation $\varphi$ is the same for all qubits at a given iteration, but varies for different iterations throughout the protocol. Here we vary it linearly, from an initial value of $\pi/4$ in the first iteration, towards zero in the last iteration.  The first iteration with $\varphi=\pi/4$ generates the uniform superposition $\ket{\psi^0} = \ket{+}^{\otimes N}$, while subsequent steps generate states that are progressively more concentrated near the best-known solutions, to better search over nearby local optima.  Similar initial states have been employed in quantum annealing algorithms \cite{duan2013alternative,grass2019quantum}.

We then apply a unitary to this state to implement approximate quantum imaginary time evolution (\ref{QITE}) by matching Pauli expectations at leading order, with details in the next section.  There is some arbitrariness in the choice of unitary.  In principle state tomography can be used to determine an exact unitary, but this is prohibitively costly in general cases \cite{motta2020determining}. A useful alternative is to employ a variational ansatz \cite{mcardle2019variational}. We choose the fully-connected pairwise ansatz
\begin{equation} \label{ansatz} U^\text{QITE}(\bm \theta) = \prod_{i<j} e^{-i \theta_{ij} G_{ij}} .\end{equation}
To approximately solve MaxCut, we choose generators
\begin{equation} G_{ij} = Z_i Y_j + Y_i Z_j \end{equation}
The ansatz terms are ordered by implementing all terms involving the first qubit, followed by all remaining terms involving the second qubit, and so on, analogous to Ref.~\cite{morris2024performant}. For $N$ qubits the circuit can be implemented in $\mathcal{O}(N)$ depth by executing gates in parallel, similar to Ref.~\cite{kivlichan2018quantum}.

This choice of ansatz and generators is motivated by previous work \cite{morris2024performant}, where Pauli $ZY$ rotations were found to be highly effective.  As discussed there, the generators can be derived as leading-order counterdiabatic terms in a suitable adiabatic evolution connecting the initial state to the ground state space of $H$.  They act on eigenstates of the $Z_i Z_j$ terms in the Hamiltonian to drive transitions between its $\pm 1$ eigensectors as
\begin{equation} (Z_i Y_j + Y_i Z_j)\frac{\ket{0_i0_j} \pm \ket{1_i1_j}}{\sqrt{2}} \propto \frac{\ket{0_i1_j} \pm \ket{1_i0_j}}{\sqrt{2}}.\end{equation}
The $\exp(-i \theta_{ij}G_{ij})$ therefore produce rotations between eigenspaces of the individual terms in $H$, to directly minimize the energy.

\subsection{Derivation of $U^\text{QITE}$} 

In standard formulations of variational quantum imaginary time evolution, the imaginary time derivatives of each side of (\ref{QITE}) are equated to each other to derive parameters that minimize the distance between the states, following the McLachlan variational principle \cite{mcardle2019variational}.  However, in previous work \cite{morris2024performant} it was found that a relaxed procedure, based on matching the Pauli expectations of the states rather than the states themselves, yielded a simpler overall procedure and gave high quality results.  We use this approach here to maintain a simple connection to Ref.~\cite{morris2024performant}; future work may consider whether better results are obtained by applying the McLachlan variational principle instead.

Using states $\ket{\psi^\text{QITE}}$ and $\ket{\psi^\text{ITE}}$ from (\ref{QITE}), we compute Taylor series of the expectations $\bra{\psi^\text{QITE}} P_{kl}\ket{\psi^\text{QITE}}$ and $\bra{\psi^\text{ITE}} P_{kl}\ket{\psi^\text{ITE}}$ for each term $P_{kl} = Z_k Z_l$ in the Hamiltonian (\ref{H}).  We truncate each series at leading order in $\tau$ to obtain a tractable small-$\tau$ approximation, with errors at $\mathcal{O}(\tau^2)$. Including higher order terms in the series would further improve the accuracy but the analysis is more difficult; we will not address it here. 

From our procedure, we obtain
\begin{equation} \label{basic QITE} 2\mathfrak{R}\sum_{i<j} \bra{\psi} P_{kl}  \frac{\partial\ket{\psi}}{\partial \theta_{ij}} \dot \theta_{ij} = - \bra{\psi} \{ P_{kl}, H-\langle H \rangle\} \ket{\psi} \end{equation}
where $\dot \theta_{ij} = \partial\theta_{ij}/\partial \tau$ and the curly brackets on the right denote the anticommutator. This is equivalent to the linear system of equations
\begin{equation} \label{QITE equations} \mathcal{G} \cdot \dot{\bm{\theta}} = \bm{\mathcal{D}} \end{equation}
with 
\begin{align} \label{G} \mathcal{G}_{(kl),(ij)} = \mathfrak{R} \bra{\psi} P_{kl}  \frac{\partial\ket{\psi}}{\partial \theta_{ij}}\\
\label{D} \mathcal{D}_{(kl)} = - \frac{1}{2}\bra{\psi} \{ P_{kl}, H-\langle H \rangle\} \ket{\psi} \end{align}
Evaluating $\mathcal{G}$ and $\bm{\mathcal{D}}$ for a given state, the linear system can be solved to determine $\dot{\bm{\theta}}$.  We then set parameters $\bm \theta = \Delta \tau \dot{\bm{\theta}}$ in $U^\text{QITE}$ to implement approximate imaginary time evolution for an imaginary time step $\Delta\tau$. 

We aim to implement the imaginary time evolution on the simple product state (\ref{superposed state}).  To do this we will derive expressions for $\mathcal{G}$ and $\bm{\mathcal{D}}$ for these states, so that we may solve the system of equations (\ref{QITE equations}) on a conventional computer, without running the quantum device.

We begin by evaluating terms in (\ref{basic QITE}) for a generic product state $\ket{\psi}$ and $\theta_{ij}= 0$ for all $i,j$, giving
\begin{equation} \mathfrak{R}\bra{\psi} P_{kl}  \frac{\partial\ket{\psi}}{\partial \theta_{ij}}   = - \frac{1}{2}\langle i[Z_kZ_l,G_{ij}]\rangle \end{equation}
Diagonal matrix elements are obtained when the qubits are identical, $(ij)=(kl)$, 
\begin{equation} \label{G1} \mathcal{G}_{(ij),(ij)} = -\langle X_i\rangle - \langle X_j\rangle.\end{equation}
Nonzero off-diagonal elements are obtained when one qubit is the same, for example $i=k$, 
\begin{equation} \mathcal{G}_{(il),(ij)} = -\langle X_i\rangle \langle Z_l\rangle  \langle Z_j\rangle\end{equation} 
When all qubits are distinct, the commutator is zero. These relations specify all elements of $\mathcal{G}$ for product states.

Next we derive terms $\mathcal{D}_{\alpha}$ for a given $\alpha=(ij)$.  We begin with the anticommutator expectation in (\ref{basic QITE}), which equates to
\begin{equation} \bra{\psi} \{ P_{kl}, H-\langle H \rangle\} \ket{\psi} = -2[\langle H \rangle \langle P_{kl}\rangle - \langle HP_{kl}\rangle] \end{equation}
For terms $P_{ij}$ in $H$ which do not act on qubits $k$ or $l$, for our product initial state the second expectation on the right $\langle P_{ij} P_{kl}\rangle$ separates to cancel with an identical expectation $\langle P_{ij}\rangle\langle P_{kl}\rangle$.  Hence, the nonzero contributions come only from terms where $(kl)$ and $(ij)$ share at least one index. Summing contributions from all such terms we obtain the elements of $\bm{\mathcal{D}}$ as

\begin{align} \mathcal{D}_{(kl)} & = \langle Z_k\rangle^2\langle Z_l\rangle^2-1 -\langle Z_k\rangle (1 - \langle Z_l\rangle^2)\sum_{i \in \mathcal{N}_l\backslash k}\langle Z_i\rangle  \nonumber\\ 
& -\langle Z_l\rangle(1  - \langle Z_k\rangle^2)  \sum_{i \in \mathcal{N}_k\backslash l}\langle Z_i\rangle  \end{align}
where $\mathcal{N}_k\backslash l$ denotes the neighborhood of vertex $k$ excluding $l$ and similarly for $\mathcal{N}_l\backslash k$. 

For our initial states, the expectation values in $\bm{\mathcal{D}}$ and $\mathcal{G}$ have the simple forms
\begin{equation} \label{Pauli expectations} \langle Z_i \rangle = (-1)^{z_i^\text{best}}\cos(2\varphi), \ \ \ \langle X_i \rangle = \sin(2\varphi). \end{equation}
Using Eqs.~(\ref{G1})-(\ref{Pauli expectations}), the $\mathcal{G}$ and $\bm{\mathcal{D}}$ are computed numerically for a given initial state (\ref{superposed state}). The parameters $\bm\theta =\bm{\dot \theta}\Delta \tau$ in  $U^\text{QITE}$ are then determined by the linear system Eq.~(\ref{QITE equations}), which can be solved in various ways \cite{press1992numerical}; we use the default solver in the \texttt{LinearSolve.jl} package \cite{LinearSolve}. The time complexity of determining $U^\text{QITE}$ is $\mathcal{O}(\text{poly}(N))$, as the $\mathcal{G}$ and $\mathcal{D}$ contain $\mathcal{O}(\text{poly}(N))$ elements each computed from $\mathcal{O}(\text{poly}(N))$ terms, while the complexity of solving the linear system of equations is polynomial in the matrix dimensions, which again are $\mathcal{O}(\text{poly}(N))$.


Using these unitaries, we evolve the state 
\begin{equation} \label{qite states} \ket{\psi} = U^\text{QITE}\ket{\psi^0}\end{equation}
at a given iteration, then sample to potentially identify a new best-known solution, following the procedure outlined in Fig.~\ref{schematic}. This process repeats for a fixed number of iterations, as described previously.  The final solution is the best observed bitstring from all samples.

\begin{figure}
    \centering
    \includegraphics[width=\linewidth]{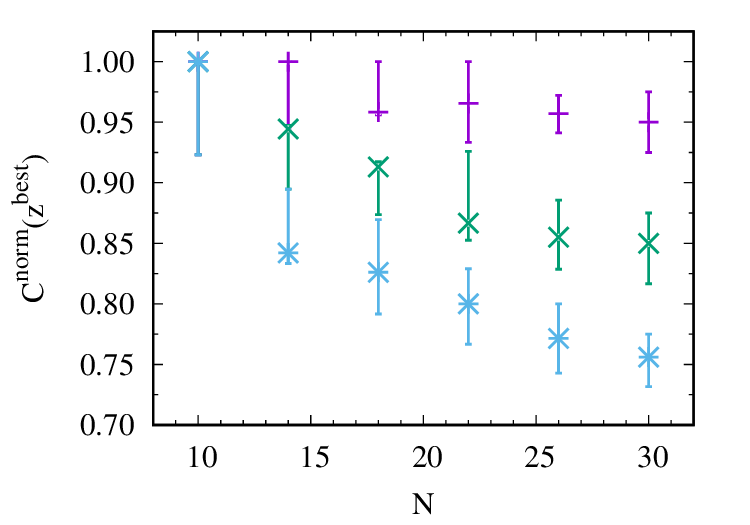}
    \includegraphics[width=\linewidth]{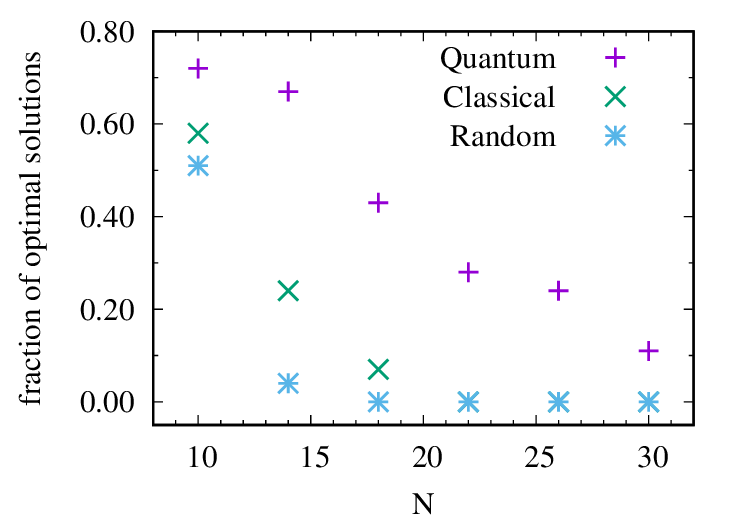}
    \caption{Performance of the protocol for MaxCut on collections random 3-regular graphs. (top) Using 100 total measurements the approach yields median normalized cost values that are close to optimal for $N\leq 30$, outperforming random search and a ``classical" version of the protocol that omits the quantum imaginary time evolution.  Error bars show quartiles of the distributions. (bottom) The algorithm obtains significant optimal solution probabilities out to $N=30$, while the classical and random searches do not. The observed fraction of optimal solution is zero for the random search at $N \geq 18$ and for the classical search at $N \geq 22$.  }
    \label{fig:performance simple protocol}
\end{figure}

\section{Simulation of the algorithm}

Here we consider numerical simulations of the approach targeting MaxCut on 3-regular graphs.  We simulated the algorithm on collections of 100 random graphs with various numbers of vertices $N$, with $N_\text{it}=10$ iterations in the algorithm, and with $N_\text{shots/it}= 10$ shots at each iteration to attempt to identify a new best-known solution that updates the initial state.  In total, this amounts to $N_\text{shots} = N_\text{shots/it} \times N_\text{it} = 100$ shots per problem instance.  The small shot budget is chosen so the algorithm operates in a limit where variational optimization would not be possible, and where random sampling would be unlikely to return an optimal solution. We used a fixed imaginary timestep $\Delta \tau=0.1$. In general we expect better results to be obtained with more shots, and that the number of shots and $\Delta \tau$ should scale with the problem size to maintain a reasonable accuracy, though we will not address the optimal scalings here.

First we evaluate the cost of the final solutions relative to the optimal ones.  We use the normalized cost 
\begin{equation} C^\text{norm}(\bm z^\text{best}) = \frac{C(\bm z^\text{best}) - C^\text{max}}{C^\text{min}-C^\text{max}} \end{equation}
where $C^\text{min}$ is the minimal (optimal) value of the cost function and $C^\text{max}$ is its maximum value.
The normalized cost is bounded as $0 \leq C^\text{norm}(\bm z^\text{best}) \leq 1$ with $C^\text{norm}(\bm z^\text{best}) = 1$ signifying the optimal solution.

The top panel of Fig.~\ref{fig:performance simple protocol} shows simulated results on the collections of graphs at various sizes $N$. The normalized costs in the top panel obtain median values that are 95\% or more of the optimal throughout the considered range of sizes.  To better assess the performance and the impact of different components of the algorithm, we compare its results against a random sampling procedure and against a similar iterative optimization using samples taken from the superposed initial state (\ref{superposed state}), denoted ``classical" in the figure since the states are unentangled. The random approach obtains normalized costs close to one at  $N=10$, but the performance drops quickly as $N$ increases. Samples from the unentangled states (\ref{superposed state}) achieve somewhat better performance, indicating that the iterative updates and sampling of the states (\ref{superposed state}) improves the performance relative to the random search.  However, the normalized cost of the final solution is consistently worse than the protocol including $U^\text{QITE}$, and decreases more rapidly with system size.

Figure \ref{fig:performance simple protocol}(bottom) shows the fraction of instances at each size for which an optimal solution was found.  The observed probability is 11\% at $N=30$, and typically higher for the smaller $N$.  This is somewhat remarkable, as the $N=30$ results have $2^{30} \approx 10^9$ possible solutions, so an extremely small optimal solution probability would be expected when randomly sampling only 100 states per instance.  Empirically, we find optimal solution fractions of zero for $N \geq 18$ using the random search and $N\geq 22$ using the classical search.  Evidently the quantum evolution $U^\text{QITE}$ significantly boosts the optimal solution probability.

For a given shot budget $N_\text{shots}$, it is interesting to consider the optimal division of shots between iterations $N_\text{it}$ and the number of shots per iteration $N_\text{shots/it}$.  A greater number of shots per iteration provides a more thorough search of nearby states at each iteration, to determine the best local optimum $\ket{\bm z^\text{best}}$ for the next step, while a greater number of iterations allows greater exploration of the space of states.  

\begin{figure}
    \centering
    \includegraphics[width=\linewidth]{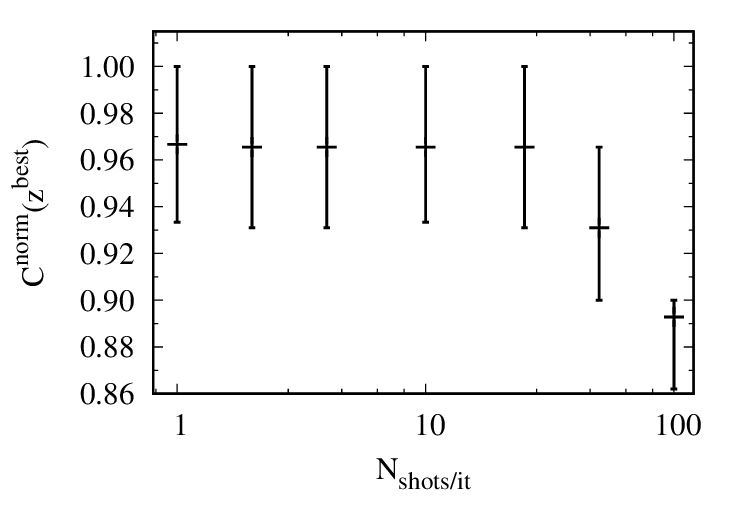}
    \includegraphics[width=\linewidth]{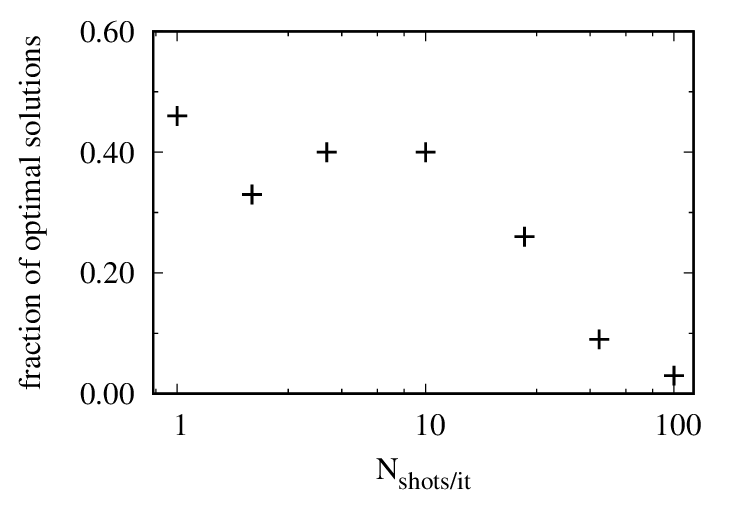}
    \caption{Performance at $N=22$ as the number of iterations $N_\text{it}$ and the number of shots per iteration $N_\text{shots/it}$ are varied, with a fixed shot budget $N_\text{shots}=100$. }
    \label{fig:vary Nmeas}
\end{figure}

In Fig.~\ref{fig:vary Nmeas} we assess the performance at $N=22$ with varying $N_\text{it}$ and $N_\text{shots/it}$, with a fixed shot budget of $N_\text{shots} = 100$.  The results appear  mostly steady for $N_\text{shots/it} \leq N_\text{it}$, while they become systematically worse when $N_\text{shots/it} > N_\text{it}$.  A likely reason for the poor behavior at $N_\text{shots/it} > N_\text{it}$ is that the algorithm does not sufficiently explore the space of states, focusing instead on sampling near suboptimal local minima that are found within the first few iterations. 

\begin{figure}
    \centering
    \includegraphics[width=\linewidth]{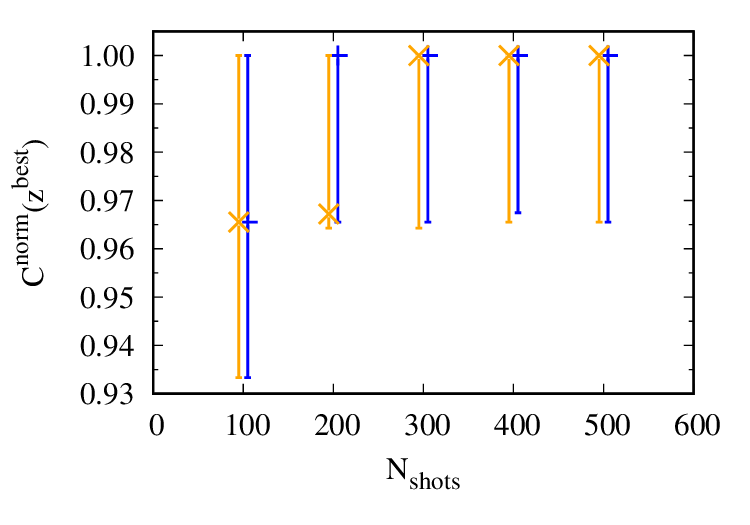}
    \includegraphics[width=\linewidth]{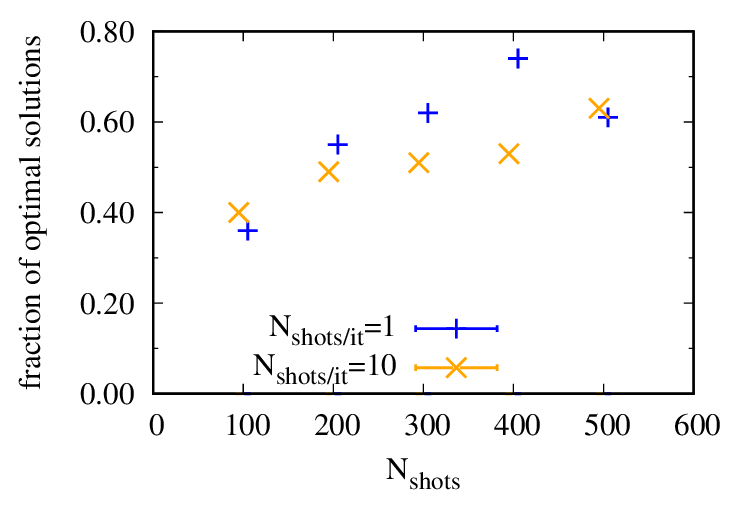}
    \caption{Performance at $N=22$ as the number of shots $N_\text{shots}$ is varied. }
    \label{fig:vary Nshots}
\end{figure}

Figure \ref{fig:vary Nshots} assesses whether the algorithm empirically approaches the optimal solution as the number of shots increases, with $N_\text{shots} \in \{100,200,300,400,500\}$. We consider cases with $N_\text{shots/it}=1$ and $N_\text{shots/it}=10$, which were found to give comparable performance in the previous figure, with offsets in the figure for visual clarity.  The algorithm succeeds in finding optimal solutions in about half or more of cases when $N_\text{shots} \gtrsim 200$, depending on the choice of $N_\text{shots/it}$.  The optimal solution probabilities appear to slowly increase towards one as additional shots are taken, with some significant fluctuations, indicating benefits of additional shots and the hoped-for approach towards optimality.

\section{Discussion} 

We presented an algorithm that uses iteratively-updated, warm-started quantum states for approximately solving combinatorial optimization problems.  The algorithm has a few desirable properties, such as employing nonvariational and shallow circuits, and produced optimal solutions with nonnegligible probability on a small shot budget.  

There are several interesting future directions for this work.  One direction is to explore higher-order variations of the quantum imaginary time evolution, which would be expected to improve the performance.  The analytic equations become much more complicated at the next order, but for the simple product initial states it may still be possible to make sense of them.  Another interesting option is to include nonlinear parameter schedules for $\varphi$ in the initial state, to prevent states from getting stuck in local minima.  Many additional variations can be made to the basic algorithmic structure of Fig.~\ref{schematic}, including different types of initial states and quantum evolutions, as well as potential connections to quantum-enhanced simulated annealing and related optimization algorithms.  Further developments along these lines are expected to enable more efficient quantum optimization algorithms for current and forthcoming quantum computers. 

\section*{Acknowledgments}

We thank Lucas Braydwood for feedback and discussions.

\bibliographystyle{unsrt}
\bibliography{references}
\end{document}